# Experimental study of developing free-falling annular flow in a large-scale vertical pipe

Yunpeng Xue[a], Colin Stewart[b], David Kelly[b], David Campbell[b] and Michael Gormley[b]

[a] Future Resilient Systems, Singapore-ETH Centre, ETH Zurich, Singapore

[b] School of Energy, Geoscience, Infrastructure and Society, Heriot-Watt University, Edinburgh, UK.

*Abstract*

Annular flow is the primary characteristic of unsteady wastewater flow, which initiates entrained air and sets up the air pressure regime within the system - an important design consideration. This paper reports on an experimental investigation of free-falling annular flow in a vertical pipe with different inlets at extended flow ranges up to Re = 3 × 10$^4$, similar to those in Building Drainage Systems (BDS). In the experimental setup, a vertical pipe system (5 m) was used to record velocity profiles and film thickness in the developing region through Particle Image Velocimetry (PIV) measurements. Entrained droplets were collected through a separator, and the entrainment fraction was calculated at different flow conditions. The study reports on the development process of the film velocity and thickness along the vertical pipe, which agrees well with empirical predictions. The results of the droplet entrainment of a vertical annular flow show the development process to the steady state. Additionally, a Tee-junction inlet in drainage system generates a higher and different entrainment profile.

## 1  Introduction

The recent COVID-19 pandemic has highlighted the importance of the building drainage system (BDS) in securing public health. Recent research has shown that pathogens such as SARS and SARS Cov-2, which cause COVID-19, as well as laboratory surrogate pathogens such as pseudomonas putida (bacterium) and PMMoV (virus), can be transmitted through the air core of the annular flow inside the BDS [1-5]. The flow regime in a drainage system is generated by the unsteady transient flow of a free-falling annular flow, which is a simplified version of a two-phase annular flow with zero or natural airflow. The unsteady nature of the flow, its large range, droplet entrainment, and the presence of solid matter highlight the complex characteristics of a real multiphase flow within a drainage system. These complexities pose limitations to the current design standards and, consequently, impact the overall quality of the system.

There has been a long and rich history of research dedicated to studying the annular flow and free-falling film in vertical pipes, owing to their significant involvement in numerous industrial processes and the complex nature of their flow mechanisms [6-9]. Vassallo [10] conducted a near-wall measurement of velocity in the liquid film using a hot-film probe and proposed a modified law of the wall for the thick film near the transition regime. Karapantsios et al. [11] successfully characterised the falling film using conductance probes. They analysed various statistical moments, such as film thickness and velocity, to gain a comprehensive understanding of the film behaviour. Hasan et al. [12] employed a multi-probe film sensor to investigate the interfacial structure of freely falling liquid films in a vertical large diameter pipe (127 mm) at Reynolds numbers ranging from 618 to 1670. In contrast to smaller diameter pipes where the waves typically exhibit a coherent ring-like pattern, the observed waves in the larger diameter pipe were localised around the circumference and exhibited axial evolution over time. Measured using a wire mesh sensor, it was found that the average liquid film at the top axial position differed significantly from those at the lower axial positions [13]. Interestingly, the data presented a distinctly different slope compared to existing correlations and theoretical models, suggesting a potential change in the film structure for large-diameter pipes.

Ho and Hummel [14] investigated the velocity distributions within falling films across a range of liquid Reynolds numbers (31-700). Their findings revealed that the fully-developed mean velocity profiles inside the films were solely dependent on the Reynolds number, irrespective of the distance from the liquid inlet. In a study comparing the measured velocity profiles of a free-falling annular flow (using the dye tracer technique) to Nusselt's predictions, good agreement was found at a low liquid Reynolds number. However, significant differences were reported in wavy turbulent films [15]. Interestingly, this work indicated, based on the measurements of the instantaneous velocity profiles inside wavy falling films, that the local laminar or turbulent character depends on the local root mean square of the film roughness, not the local Reynolds number. Adomeit and Renz [16] observed that the flow in the wave crest was decelerated as its momentum was partially transferred into the near-wall region. They found that at a Reynolds number of 200, the wave shapes became highly unsteady, and approximately every second wave collision led to the formation of turbulent spots.

More recently, researchers have employed advanced techniques such as simultaneous planar laser-induced fluorescence and a combination of particle image and particle tracking velocimetry [17, 18] to obtain detailed velocity profiles in annular flow. These studies have provided valuable insights into the local and instantaneous velocity fields beneath the interfacial waves, as well as the mean velocity, velocity fluctuation, and kinetic energy profiles within the liquid film. The planar laser-induced fluorescence (PLIF) method has demonstrated promising results due to its relatively high temporal and spatial resolution [19-22], coupled with an improved visualisation algorithm [23]. The measured average film thickness data exhibited good agreement with previous experimental data and was comparable to Nusselt's theory [18] at low Reynolds numbers. However, with increasing Reynolds number, the film thickness was increasingly underpredicted by the theory, but with good agreement with Mudawwar and El-Masri's semi-empirical turbulence model [24].

The formation of droplets and their entrainment in gas-liquid two-phase flow are caused by surface instability on the film due to the relative velocity between the film and gas, which has been extensively discussed and summarised by Azzopardi [9] and Berna et al. [7, 8]. However, in the case of free-falling film flow in a vertical pipe, the velocity difference between the liquid film and the gas phase is usually not significant enough to cause wave detachment and droplet generation. Instead, the entrainment of liquid droplets occurs as a result of the natural formation and breakup of waves travelling on the interface of the falling film. This phenomenon is strongly influenced by the properties of the liquid, particularly surface tension and viscosity.

Droplet entrainment plays a crucial role in drainage systems, especially in high-rise buildings. Due to the significant velocity difference between the entrained droplets and the annular film, the droplets travel at a much faster rate, and the film flows downward along the pipe after the drops. This highlights the need for a deeper understanding of the complex mechanisms involved in free-falling film flow by capturing detailed information about the flow characteristics. However, only a few studies have focused on investigating the entrainment of liquid droplets in a free-falling annular flow in vertical pipes [25] and the droplet entrainment mechanism in a falling film annular flow with zero or natural airflow remains unclear.

Therefore, the current study aims to enhance the comprehension of the flow development process and droplet entrainment in annular flow with natural or negligible airflow, specifically examining the influence of different inlets, particularly the Tee junction of a drainage stack. To achieve this, particle image velocimetry is utilized to capture the flow characteristics of a free-falling annular flow in a large-diameter vertical pipe at high Reynolds numbers. The study measures and compares the film velocity, thickness and droplet entrainment at various locations with previous experimental data and prediction models.

## 2 Experimental configurations and procedures
### 2.1 Experimental setup

The experiments were conducted in the drainage research lab, at Heriot-Watt University, where the pipeline was built. The flow configuration is presented in Figure 1 (a), which contains (1) a $3 \times 2 \times 1.8$ m³ water reservoir, (2) a centrifugal pump up to the liquid flowrate of Q =10 L/s, (3) a turbine flow meter with a measurement accuracy of 0.5%, (4) an air ventilation pipe (ID =102 mm), (5) a hot-wire anemometer (AVUL-5DA1) with a reading accuracy of 5%, (6) annular flow inlet or 'Tee' junction inlet, (7) test section for PIV or droplet entrainment measurement, and (8) replaceable pipe to allow the measurement at different heights. In the experiment, water was pumped up using a centrifugal pump and sent into either the annular inlet or a standard 110 mm 'Tee' junction after passing through a turbine flow meter. Air was drawn into the vertical pipe from the horizontal ventilation pipe and filled the core region of the annular flow. The air anemometer was mounted on the horizontal pipe to collect the airflow velocity during the experiment. The present study conducted a test campaign at high liquid Reynolds numbers, ranging from 2900 to 45000. The liquid Reynolds number is defined as $Re_l = j_l D g / \mu_l$, where $j_l$ is the liquid superficial velocity, $D$ is the inner diameter of the vertical pipe, and $\mu_l$ is the liquid dynamic viscosity.

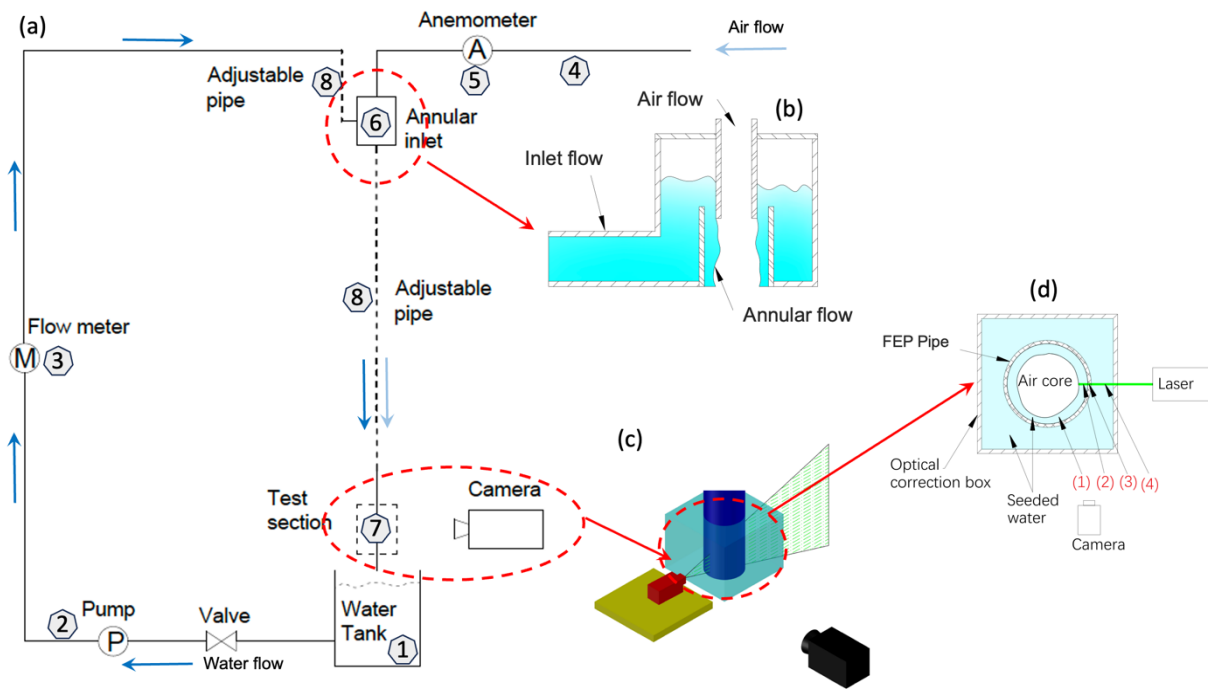

Figure 1. A simplified sketch of the pipeline system used in the current work (a), a detailed configuration of the annular inlet (b), PIV setup with an optical correction box in the test section and (c), and cross-section view of the test section showing the optical correction box, FEP pipe, annular flow and air core (d).

The design of the annular inlet in our experiment was meticulously developed to induce a stable annular flow configuration, as illustrated in Figure 1 (b). At the base of the inlet, a diameter of 102 mm was established, aligning with the inner diameter of the principal vertical PVC pipe. A short pipe with an external diameter of 80 mm was introduced into this cavity, enabling a maximum film thickness of 11 mm. To replicate the flow within a drainage stack, a standard 110 mm 'T' junction (with an internal diameter of 102 mm) was incorporated. Both inlets, characterised by their adaptable height and an outside diameter of 110 mm, could be seamlessly connected to the transparent vertical PVC pipe (depicted in blue). This strategic arrangement offered the flexibility to modify the measurement location through adjustments in the pipe's length and the inlets' positioning. Throughout this study, measurements were undertaken at four specific locations, each positioned at a specific distance below the inlet. These locations were 5, 15, 25, and 40 diameters below the inlet, represented as L/D ratios of 5, 15, 25, and 40. This carefully devised setup allowed for a comprehensive investigation of flow behaviours at various distances downstream from the inlet.

Throughout PIV measurements, a transparent optical correction box filled with water was judiciously incorporated, as showcased in Figure 1 (c, d). Within this designated test section, a relatively short (500 mm) pipe fabricated from fluorinated ethylene propylene (FEP) and boasting an inner diameter of 102 mm was employed. This choice was deliberate, as it closely aligned with the refractive index of water. By doing so, we effectively minimised optical distortions stemming from pipe curvature and refractive index disparities between the PVC pipe and the surrounding water medium.

To facilitate the imaging process, a CCD camera with a resolution of 2048 × 2048 pixels² was positioned on the opposing side of the optical correction box. For illumination, a Litron 15 Hz Nd-YAG laser emitting light with a wavelength of 532 nm was harnessed. To enable accurate flow visualisation, tracer particles were introduced into the water flow via 10-micron hollow glass beads. This imparted an estimated particle density spanning between 40 and 60 particles within the confines of the integration window. The velocity fields inherent to the PIV images were subsequently derived using the resources provided by Dantec Dynamics. Post-processing entailed the application of cross-correlation, with an integration window dimension measuring 32 × 32 pixels². With the aim of robust statistical analysis, a total of 1500 pairs of images were captured at a frequency of 10 Hz for each test configuration. The uncertainty associated with the acquired velocity fields was rigorously estimated to be within the realm of $10^{-3}$ m/s, using integration window-based statistics with sub-pixel accuracy at 1/10 pixel.

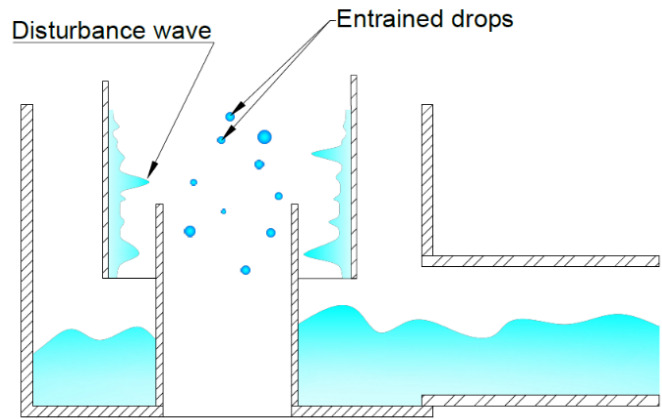

*Figure 2. The structure used to separate the entrained droplets and the annular film for entrainment measurement.*

Following the completion of PIV measurements, a specialised mechanical arrangement (as depicted in Figure 2) was employed to effectively segregate entrained droplets from the annular film, guided by the insights presented in references [20, 26]. Drawing from the visualised film, a narrower pipe with an external diameter of 90 mm was introduced into the primary pipe. This insertion introduced a 6 mm gap, facilitating the separation of droplets from the liquid film. The film was directed toward the main water tank, while beneath the separator, a cylindrical container with an inner diameter of 300 mm was equipped with three pressure sensors at its base to facilitate the collection of entrained droplets. The recorded pressure profile, mirroring the water depth, correlates directly with the volume of entrained water droplets. Subsequent refinement, aimed at mitigating the impact of surface waves, is applied to this profile. Subsequently, the fraction of droplet entrainment is computed. The duration of a single measurement varies from 0.4 to 30 minutes, contingent upon the specific entrainment rate and the volume of the container.

## 2.2   Image processing

Figure 3 offers a representative depiction of the measurement plane, illuminated by a 1 mm laser sheet, a configuration illustrated in Figure 1 (d). The composition of the image is delineated into distinct segments labelled as parts 1 to 4, each corresponding to a specific element. These are the annular film in frontal view, the tracer particle-laden annular film bathed in laser light, the 4 mm thick wall of the transparent tube, and the seeded water within the optical correction box utilized as a reference.

The utilization of an edge detection algorithm, grounded in intensity fluctuations across the image, enables the differentiation between the liquid film and the encased air core. An illustrative instance is presented in Fig. 3 (a), showcasing an amplified view of the descending film. By selecting optimal thresholds, it becomes feasible to delineate the wall of the pipe, the boundary demarcating the illuminated film and the air core, and the film as perceived from the frontal view, as portrayed in (b). Notably, the delineated film as depicted in (c) is then harnessed to isolate the effective velocity field, as demonstrated in (d).

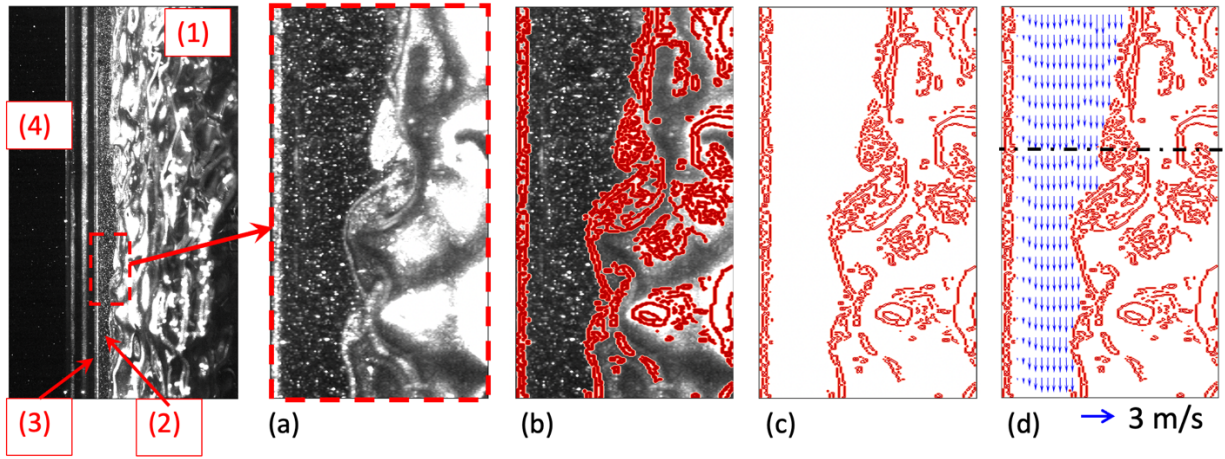

Figure 3. A typical raw image captured in the PIV measurement shows the focused annular film (left) and the gas-liquid interface detection process of the zoomed-in region of the annular film (a-d).

It is crucial to underscore that optical measurements of film thickness are susceptible to various challenges, encompassing fluctuations in film thickness, distortions in optical paths, and the occurrence of total internal reflection at interfaces. Notably, the phenomenon of total internal reflection at the gas/liquid interface stands as a significant hurdle in optical film thickness measurements, often leading to overestimations of up to 30% [19]. While the combination of large pipe diameter and the orthogonal alignment between the camera and the laser sheet can enhance the precision of film thickness estimation, the inherent uncertainty stemming from these issues remains non-negligible. Consequently, in the context of our present work, direct collection of film thickness data from optical records is not pursued due to these uncertainties.

Nonetheless, the film thickness information remains integral to our investigation into film velocity profiles. Notable examples in related research fields showcase innovative strategies to determine the gas-liquid interface from recorded images through the analysis of light intensity variations [27]. In the realm of PIV measurements, the gas/liquid interface has been discerned through the utilization of certain particles with strong blurring characteristics within the images [16, 28], incorporation of techniques from PLIF measurements [17, 29], or even deduced from the velocity profile of the liquid flow, which is quite effective in interface detection [30].

As depicted in Figure 3, our study benefits from the distinct visibility of both the seeded falling film and the film as observed from the frontal view. This distinct differentiation, facilitated by strong reflection leading to heightened intensity, serves as a pivotal factor. The intensity variation, coupled with the radial direction velocity profile, collectively contributes to the determination of the effective velocity distribution within the descending film.

## 2.3 Determination of the average film velocity

Figure 4 effectively portrays the radial distribution of velocity and intensity at the location designated by the dashed line in Figure 3 (d). The presented velocity distribution within the annular film demonstrates a good qualitative alignment with findings from prior PIV measurements [15-17]. This alignment underscores the capturing of boundary layer flows adjacent to the wall, coupled with the steady velocity observed farther away from the wall surface within the liquid medium. A pronounced velocity drop, a recurring observation across numerous images, at approximately 2.8 mm from the wall, signifies a significant feature. Importantly, this decline doesn't pertain to the velocity of the illuminated film near the interface but rather signifies the velocity attributed to the film captured in the frontal view (part 1 in Figure 3). Concurrently, the image intensity at this same location undergoes a rapid augmentation, culminating in a prominent peak around the 3.5 mm mark from the wall, as evidenced in Figure 3 (d). This conspicuous change serves as a clear indicator of robust reflection emerging from the air/water interface.

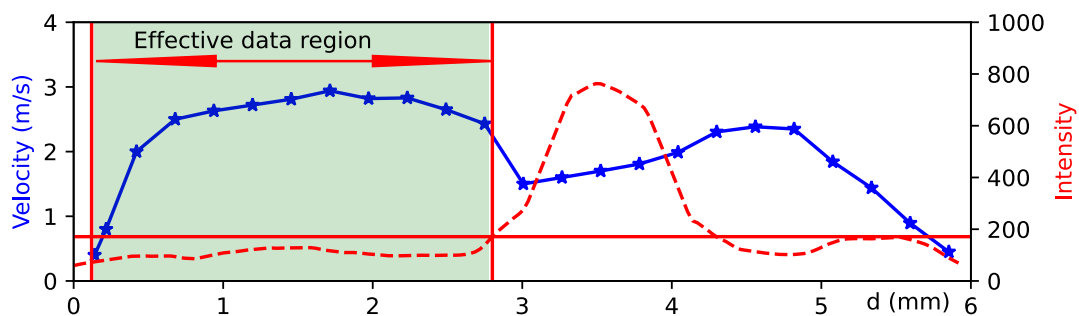

Figure 4. The velocity and image intensity profiles in the radial direction from the wall.

We employ two distinct thresholds to effectively identify the pertinent velocity profile. These thresholds are predicated on the gradients of both intensity and velocity distribution. We opt to focus on the velocity profile extending from the pipe's wall to the point of the first pronounced increase in intensity or the initial sharp drop in velocity, whichever occurs first. This carefully selected range defines the effective velocity profile, which in turn enables us to ascertain the spatially averaged velocity. To illustrate this process, consider the case of Figure 4. Within this context, the instantaneous spatially averaged velocity is quantified as 2.29 m/s.

Displayed in Figure 5 (left) is the evolution of the spatially averaged film velocity (at L/D = 25, Q = 1.42 L/s). The plot effectively portrays the velocity's fluctuation, oscillating within the vicinity of 2 m/s. This fluctuation manifests an upper bound at 3.3 m/s and a lower limit of 0.6 m/s. Notably, it's important to underscore that this representation serves as a statistical summary of the average velocity, capturing the broader trend rather than the intricacies of the velocity's temporal variation. The limited sampling frequency, in comparison to the rapid flow fluctuations, precludes us from obtaining a time-series velocity profile.

It is pertinent to acknowledge that the velocity distribution within the descending film is characterised by distinct patterns: a boundary layer flow adjacent to the wall and a more stable velocity further within the film. Given this unique velocity distribution, the computation of average velocity derived from the radial velocity profile is appreciably robust against potential errors in film thickness estimation. For instance, even under the most adverse scenario involving a 30% overestimation of film thickness, the average velocity computed within the 0 to 2 mm range is determined to be 2.18 m/s, reflecting a negligible 2.14% error. Underpinning this analysis, the comprehensive uncertainty evaluation, anchored in the most unfavourable scenario, registers at 2.48%, as highlighted in Figure 5 (right). Among the various measurement conditions, the maximum average uncertainty is observed at L/D = 15 and Q = 0.466 L/s, culminating at 5.62%. Importantly, all data points corresponding to film thickness smaller than 1 mm, which is selected based on the minimum thickness of 1.51 mm in Figure 7 below, are excluded from the velocity averaging process and error estimation, ensuring the robustness and accuracy of the calculation. Utilizing both the liquid flow rate and the computed average velocity, we are able to deduce the average film thickness. A summary of the uncertainties associated with the measurements conducted in this study has been thoughtfully compiled and is presented in Table 1.

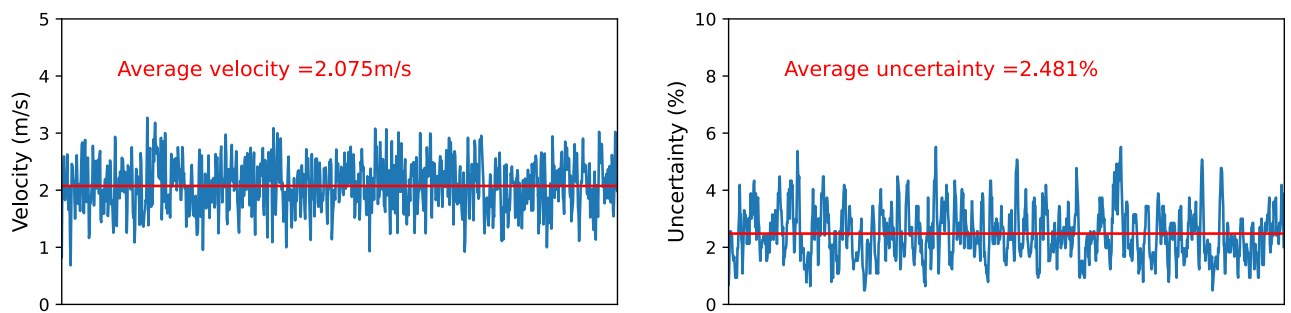

Figure 5. An example of the spatially averaged film velocity and uncertainty based on the worst scenario of a 30% overestimation of the film thickness against time.

Table 1. Estimated overall uncertainties of the focused variables in the current work

| Variables | Uncertainty |
|---|---|
| Liquid flow rate | 0.5% |
| Air velocity | 5% |
| Averaged film velocity | 5.62% |
| Film thickness | 5.64% |
| Entrainment | 2.3% |
| Entrainment fraction | 2.5% |

## 3 The falling film along the pipe

### 3.1 Velocity profile

In Figure 6, the average film velocity is depicted as a function of the liquid flow rate at different positions along the flow path starting from the inlet. It can be observed that as the inlet flow rate increases, the film velocity also increases at all measurement locations. For example, at a location of L/D = 5, the film velocity gradually increases from 0.63 m/s to 2.23 m/s as the liquid flow rate increases from 0.466 L/s to 2.594 L/s. At a given flow rate, the annular film is accelerated by gravity, resulting in an increasing velocity along the pipe. At an inlet flow rate of 1.42 L/s, the film velocity at L/D = 5 is 1.74 m/s, and it gradually increases to 1.91 m/s, 2.44 m/s, and 2.86 m/s at L/D = 15, 25, and 40, respectively. Therefore, it can be predicted that further downstream of the pipe (L/D > 40), the film velocity will continue to increase until it reaches the terminal velocity. In a free-falling annular flow, the forces acting on the water film are the frictions (pipe wall/water and water/air) and gravity, which lead to a terminal velocity of the water film ($V_t$) [31] as:

$$V_t = \sqrt[3]{(2gQ)/(f\pi D)} \qquad (1)$$

where $g$ is the acceleration, $Q$ is the liquid flow rate, $f$ is the friction factor of a smooth PVC pipe (~0.004), and $D$ is the diameter of the stack. The predicted terminal velocity in a 102 mm plastic pipe with a friction factor of 0.004 is also plotted in the figure, indicating that as the annular film moves downstream, it will approach a fully developed state and reach the terminal velocity at further downstream locations of L/D > 40. The flow development at high Reynolds numbers, characterised by an increasing film velocity, leading to the attainment of a steady terminal velocity downstream of the pipe, has been observed in a rectangular vertical pipe [32].

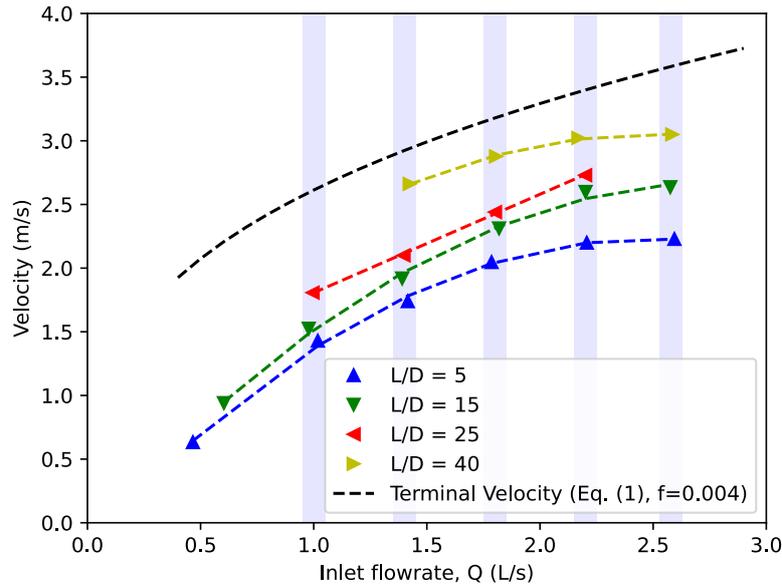

*Figure 6. Average film velocity along the vertical pipe as a function of the inlet flow rate.*

## 3.2 Film thickness

In Figure 7, determined from the flow rate and film velocity, the average film thickness at different measurement locations is plotted against the inlet liquid flow rate showing its ranges from 1.51 to 3.52 mm in this study. As the inlet flow rate increases, the film thickness also increases at all measurement locations. However, for a given flow condition, the film thickness decreases along the pipe as it accelerates due to gravity. These findings provide important insights into the behaviour of annular flows and can be used to design effective film separators that can handle the varying film thickness and fluctuation.

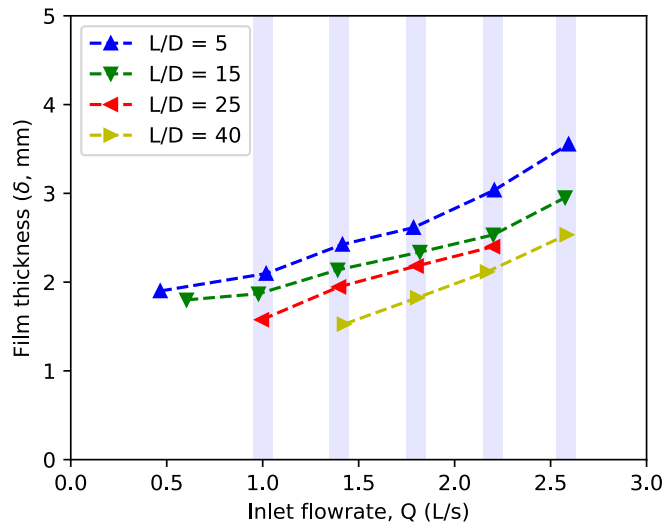

*Figure 7. Time-averaged film thickness along the vertical pipe as a function of the inlet flow rate*

It is worth mentioning that lower flow rates were also examined during the experimentation. However, due to the presence of a thin film at these lower flow rates, and an even thinner film downstream of the pipe at the same inlet flow rate, as shown in Figure 7, effective PIV measurements could not be obtained. This explains the lack of measurement results at low flow rates. Furthermore, it was observed that a significantly higher minimum flow rate is necessary to achieve reliable PIV measurements at locations further downstream (L/D = 40) compared to near the inlet (L/D = 5).

The behaviour of free-falling film along the longitudinal direction has been extensively studied in previous research [11, 12]. At low Reynolds numbers (200), the film thickness exhibits a slight decrease along the pipe. As the Reynolds number increases to 1000, there is a rapid increase in film thickness in the entrance region, which then becomes approximately constant at greater longitudinal distances [33]. In the case of high Reynolds numbers, the film becomes thinner along the longitudinal direction as a result of further flow development [34, 35]. In studies involving square pipes with Reynolds numbers exceeding $10^4$, similar to the current work, it was observed that the falling film thickness sharply increases near the entrance and then decreases along the longitudinal direction until reaching an approximately constant thickness [35]. These observations align well with the findings of the present study, where the film thickness initially increases to a maximum at the entrance and subsequently decreases as the flow further develops. Therefore, a thick falling film is observed near L/D = 5, which then gradually becomes thinner along the longitudinal direction until reaching a steady and fully developed state.

The empirical correlation generally used to predict the average film thickness of a free-falling annular flow is [13, 36, 37]:

$$\bar{\delta} = aRe_l^b \left(\frac{v_l^2}{g}\right)^{\frac{1}{3}} \qquad (2)$$

Here, $v_l$ is the liquid kinematic viscosity ($10^{-6}$ m$^2$/s), $g$ is the acceleration due to gravity (9.81 m/s$^2$), $a$ and $b$ are empirical factors. To cover a wide range of flow conditions, a comprehensive analysis was conducted by combining previous experimental data obtained from the developed flow region with the current results. These data [11, 12, 15, 18, 33, 36-41] were then compared with Nusselt's theoretical solution and two recently proposed empirical predictions [36, 37]. The comparison, as shown in Figure 8, reveals a high level of consistency among all the collected data. In the laminar region (low Reynolds number), the results align well with Nusselt's solution. As the flow transitions into the turbulent region, the two recent correlations demonstrate good accuracy in predicting the average film thickness. At high Reynolds numbers, there is a noticeable agreement in the film thickness measurements of the current work at the downstream section of the pipe. Furthermore, it indicates a consistent flow development along the length of the pipe, which is reflected in the observed film thickness variations at different measurement locations. Although not significantly pronounced, a slight discrepancy in the film thickness reported in Hasan's and Takaki's studies is observed. These measurements were taken from a relatively upstream location, suggesting that the annular film may not have fully developed at that specific measurement point. This observation aligns with our current findings of the flow development process along the pipe.

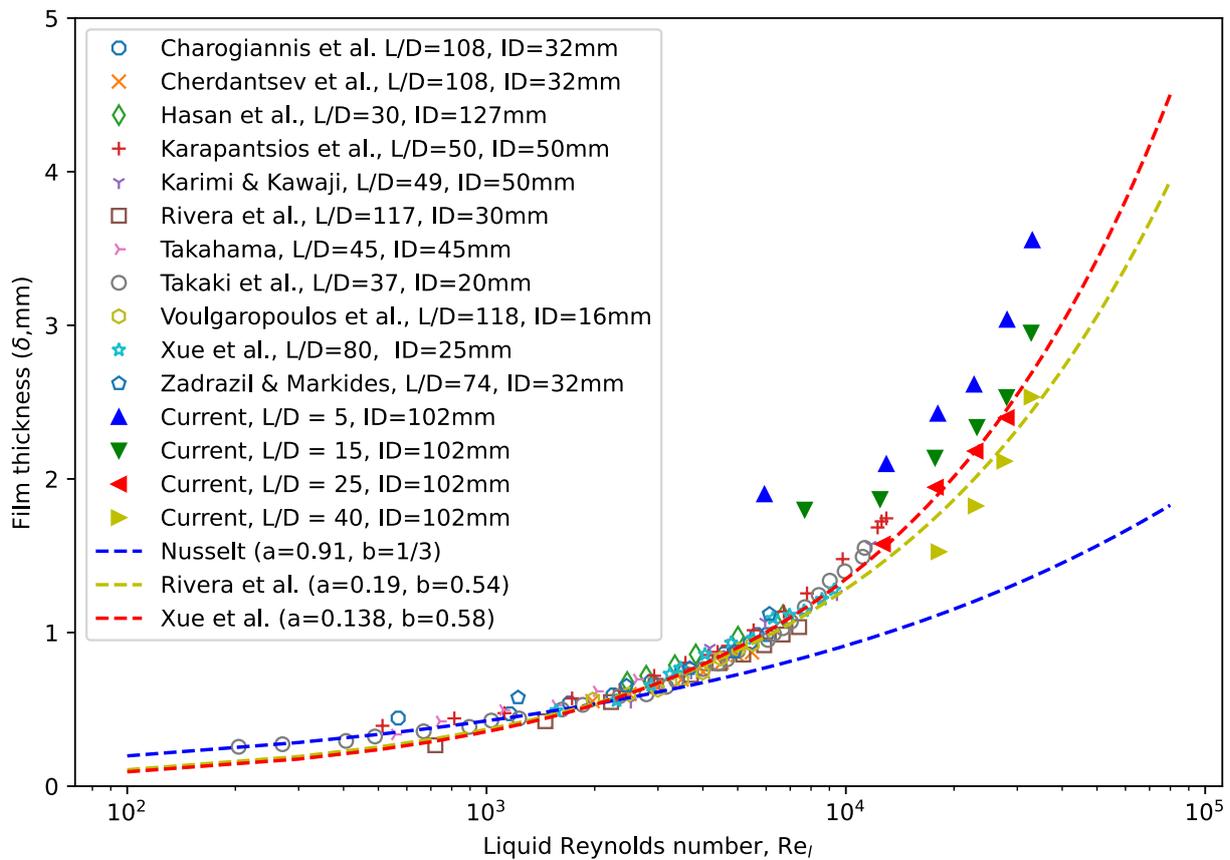

*Figure 8. Comparison of current film thickness profile with previous experimental data and empirical predictions.*

The optical measurements were also conducted using the 'T' junction inlet, which is commonly employed in drainage systems. In a two-phase annular flow, the liquid film exhibits three-dimensional structures with significant height fluctuations in the circumferential direction [36, 42-46]. In the case of a vertical pipe with a 'T' junction inlet, the unsymmetrical distribution of the film becomes more pronounced due to the radial velocity induced by the horizontal component of the inlet flow. This asymmetry becomes more prominent at higher Reynolds numbers and closer to the inlet. Consequently, a very thin film or even a dry area can be observed at the back side of the injection point, while a thicker film is present on the opposite side of the inlet. The presence of a thin film poses a challenge for optical measurements, as it restricts the feasibility of obtaining measurements in those regions. Additionally, due to the unsymmetrical nature of the flow in the circumferential direction, it becomes difficult to estimate the average velocity or film thickness accurately. These aspects are primarily influenced by the longitudinal location within the pipe and the Reynolds number at the inlet. To obtain more detailed measurements of the flow properties, the use of a multi-probe sensor with sensors arranged in both the axial and circumferential directions could be considered [12, 46].

## 4  Droplet entrainment of the free-falling flow

### 4.1  Annular inlet

The droplet entrainment fraction rate is a critical parameter for describing the amount of entrained liquid in the gas core of a free-falling annular flow in a drainage stack. This rate ($E$) is defined as the ratio of the mass flow rate of the entrained droplets to the total liquid mass flow rate and is typically measured by separating the liquid film from the entrained drops [20]. In this work, the droplet entrainment in the gas core is evaluated by analysing the pressure profiles obtained from the droplet collector.

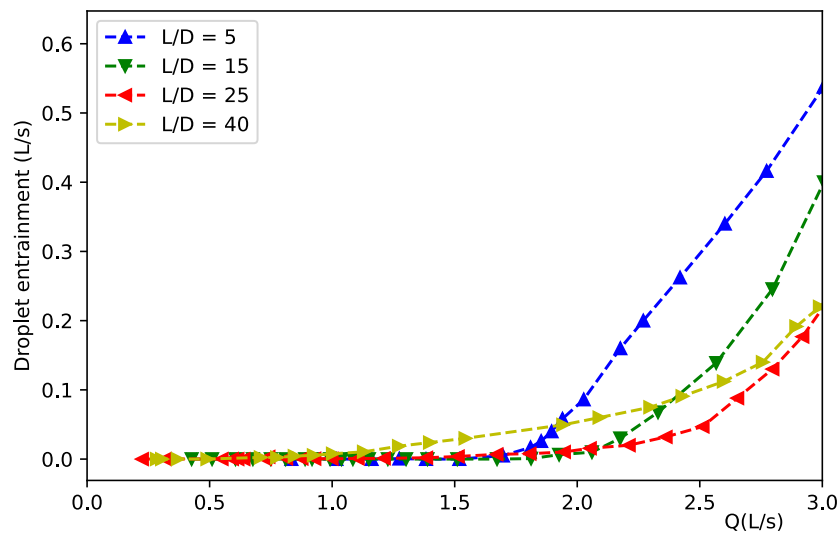

*Figure 9. Droplet entrainment against the inlet flow rate at different measurement locations below the annular inlet*

Figure 9 illustrates the droplet entrainment in the gas core of a typical annular free-falling flow at different longitudinal locations. At low inlet flow rates (Q < 1.8 L/s), there is almost negligible entrainment of liquid drops in the gas core near the annular inlet (L/D < 25). However, as the flow develops at L/D = 40, more drops are generated from wave breakup and entrained in the gas core, as evident from the entrainment profile. For example, at an inlet flow rate of 0.71 L/s, droplets can be observed in the gas core at L/D = 40 with an entrainment rate of 0.005 L/s. As the inlet flow rate increases, the presence of droplets in the gas core becomes more prominent. At an inlet flow rate of approximately 2.5 L/s, the entrainment rates along the pipe are 0.31, 0.15, 0.06, and 0.09 L/s, respectively. However, it should be noted that the deposition of entrained droplets becomes more significant in the downward annular flow, resulting in a decrease in droplet entrainment at lower measurement locations (15D & 25D). Similar findings were reported in [25], where it was observed that as the inlet Reynolds number increased, the entrainment rate gradually increased to 0.25 L/s in the region where the falling film was well developed. This observation highlights the development process of droplet entrainment and deposition in a vertical annular flow.

It is worth mentioning that the measurement quality of droplet entrainment at high Reynolds numbers is limited. This is primarily because at high Reynolds numbers, the falling film is thick, and the fluctuating wave thickness can exceed 6 mm, leading to the collection of these waves as entrained drops. Consequently, this can result in an overestimation of the droplet entrainment near the inlet (L/D=5) and at high Reynolds numbers (Re > $3 \times 10^4$). To address this challenge, the use of a droplet collector with an adjustable opening or together with other measurement devices, such as wire mesh sensors, is recommended.

The velocity of the airflow sucked from the ventilation pipe by the liquid film is measured using the hot-wire anemometer and presented in Figure 10. It is shown that the air velocity can be expressed as a function of the inlet flow rate with a good correlation.

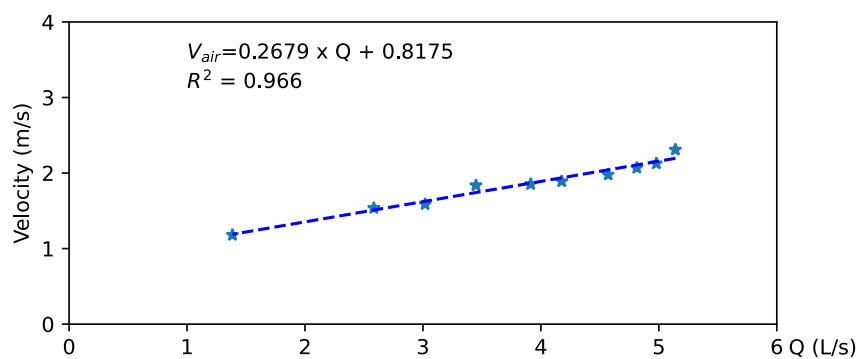

Figure 10. The average velocity of the air core as a function of the inlet flow rate.

Several theoretical and empirical correlations of the droplet entrainment fraction rate in two-phase annular flow have been proposed as a function of inlet conditions and pipe configuration [47-50]. For instance, one of the most used correlations developed by Ishii and Mishima [50] is:

$$E = tanh(7.25 \times 10^{-7} We_{air}^{1.25} Re_l^{0.25}) \quad (3)$$

$$We_{air} = \frac{\rho_{air} j_{air}^2 D}{\sigma} \left(\frac{\rho_l - \rho_{air}}{\rho_{air}}\right)^{\frac{1}{3}} \quad (4)$$

Here, $We_{air}$ is the Weber number of air core, $\rho_{air}$ and $j_{air}$ are the density and superficial velocity of the airflow, and $\sigma$ is surface tension. The experimental results are plotted against the liquid Reynolds number and compared with the empirical correlation, as shown in Figure 11. Similar to the observations of film velocity and thickness in Figures 6 and 7, the entrainment fraction rates along the longitudinal direction also indicate the flow development. While a good agreement between the experimental results and the prediction can be seen at low Reynolds numbers (Re < 1 × 10⁴), it is important to note that the negligible entrainment rate at these conditions is not sufficient to conclude that the current correlation is accurate for free-falling annular flow.

It is important to note that the existing correlations for two-phase annular flow have been developed primarily based on scenarios with a significant gas flow, where the gas superficial velocity is considered a critical parameter. However, in the context of falling film annular flow with minimal or negligible gas velocity difference, droplet formation may occur solely due to the natural formation and breakup of waves. As a result, the gas velocity or Weber number may have a negligible impact on the entrainment process. In some prior studies exploring two-phase annular flow [18, 37], the gas input was intentionally set to zero to simulate the free-falling film annular flow. While this represents a similar scenario to our current study, where air is drawn into the air core, it is essential to recognise that there might still be variations in flow properties due to the presence of the air core. The impact of the air core on droplet entrainment will be further discussed in the subsequent sections of this study.

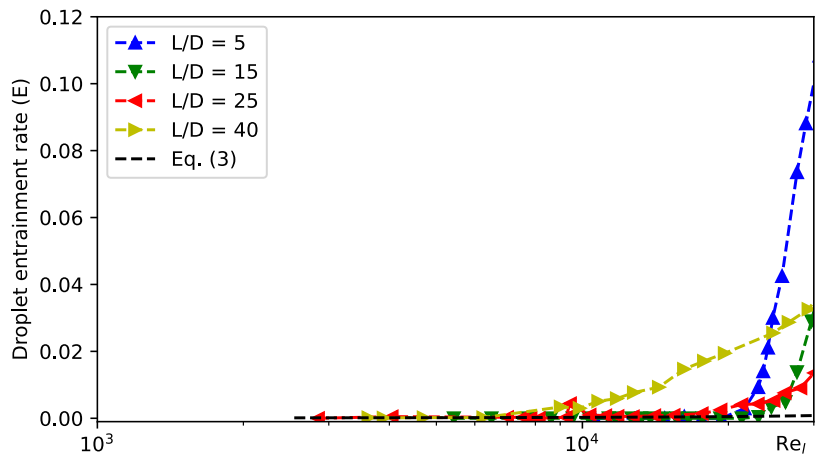

*Figure 11. Droplet entrainment rate as a function of the inlet Reynolds number compared with the empirical correlation.*

## 4.2 Tee junction inlet

When liquid is injected horizontally into a vertical pipe, for example from a 'T' junction of a drainage system, annular flow still prevails but with increased turbulence induced by the junction. Figure 12 illustrates the relationship between droplet entrainment and the liquid Reynolds number at various measurement locations. Notably, two distinct peaks are observed regardless of the measurement location, and these peaks are attributed to the horizontal injection into the vertical pipe from the 'T' junction. In this scenario, the formation of droplets differs from that of falling film annular flow. At low inlet velocities, the liquid adheres to the pipe wall due to surface tension, resulting in minimal entrainment of droplets in the air core and negligible droplet entrainment overall. However, as the inlet velocity increases, the falling liquid generates more droplets that are entrained in the air core. The first peak in droplet entrainment occurs at a downstream location around $L/D = 40$ and a Reynolds number of approximately 5500. Subsequent increases in the injected horizontal fluid lead to the first peak of droplet entrainment occurring at an earlier location, specifically at $L/D = 5$, with an entrainment rate of 0.3.

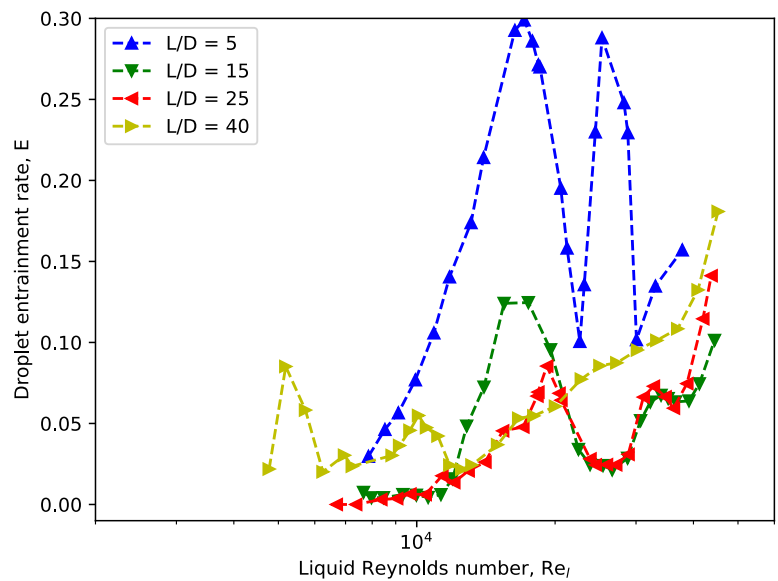

*Figure 12. Entrainment rate of an annular flow in a vertical pipe with 'T' junction against the inlet Reynolds number*

As the inlet velocity further increases, the falling liquid in the central region approaches the opposite wall tangentially, resulting in a reduction of droplets entrained in the air core. This explains the decrease in the entrainment rate after the first peak. If the liquid is injected at even higher velocities, the stream begins to impact the opposite wall, generating more droplets and causing the entrainment rate to increase again until the second peak. At higher Reynolds numbers, annular flow occurs near the junction after the stream hits the opposite wall and reverses its direction, leading to a second drop in the entrainment rate, particularly at L/D = 5. However, the second peak is less significant at further downstream locations as annular flow occurs earlier in the downward flow. The observed entrainment profile in a large-scale vertical pipe with a tangential inlet, such as a main drainage stack connected to 'T' junctions, reveals the complex mechanisms involved. Further increases in the inlet velocity result in higher droplet entrainment rates, driven by the same mechanisms as in ideal annular flow.

To investigate the impact of airflow in the core region on droplet entrainment, the air ventilation pipe depicted in Figure 2 was completely closed, resulting in zero airflows within the annular flow. The droplet entrainment rates at L/D = 25 for different configurations are shown in Figure 13. In the case of free-falling annular flow, the presence of air ventilation has a negligible effect on droplet entrainment when the inlet Reynolds number is below 30000 (inlet flow rate, Q < 2.5 L/s). Only as the inlet fluid further increases does the influence of airflow become significant, with a maximum increase of 30% observed at Re = 45000.

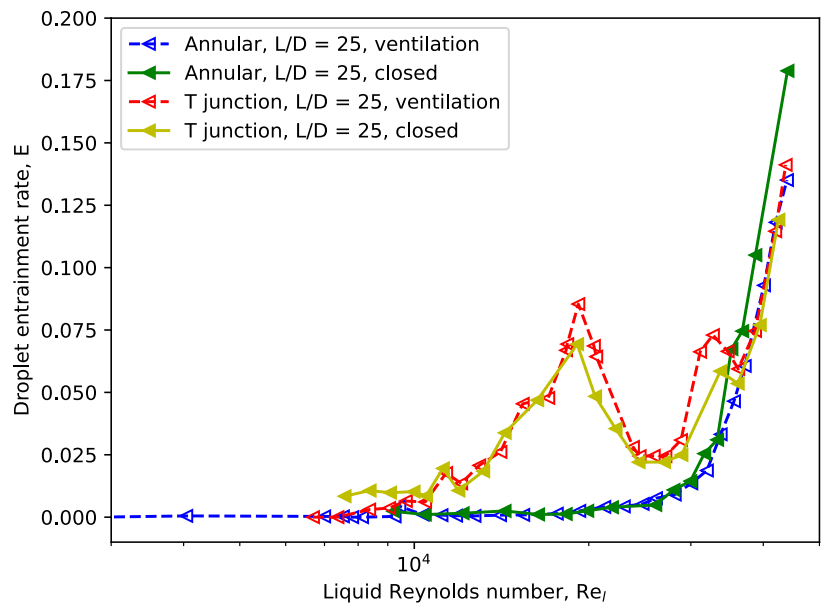

*Figure 13. Entrainment rate in the vertical pipe with and without air ventilation*

When the Tee-junction is used as the inlet, closing the air core flow results in a significant increase in droplet entrainment at a lower Reynolds number (Re < 10000). In both conditions, i.e., annular flow with Re > 30000 and Tee-junction with Re < 10000, the increase in entrainment is attributed to the lower pressure in the central region when the air ventilation is closed. The lower pressure within the air core has a positive impact on droplet atomization, leading to a higher entrainment rate. The slightly lower entrainment rate observed at the two peaks with the Tee-junction configuration is attributed to reduced turbulence of the stream in the air core when the airflow is stopped. Based on these findings, it can be concluded that the presence of airflow in the central region is not a prerequisite for droplet entrainment. Instead, the pressure difference between the inlet liquid and the central air core plays a significant role in droplet entrainment.

## 5 Conclusion and remarks

In conclusion, this study presents a comprehensive experimental investigation of free-falling annular two-phase flow in a vertical pipe with different inlet configurations. By measuring film velocity, film thickness, and droplet entrainment in the air core, we have gained valuable insights into the flow development process and highlighted the need for further research on unsteady flow behaviour, especially from inception to steady state.

The results clearly demonstrate that the averaged film velocity increases with the inlet flow rate, leading to a larger falling velocity of the annular film. Along the vertical pipe, the average flow velocity approaches the terminal velocity of a falling film further downstream, revealing a consistent trend in the flow development process. The averaged film thickness data, obtained from flow rate and film velocity, exhibit good agreement with previous experimental data and empirical predictions, even at extended flow ranges up to $Re = 3 \times 10^4$. This confirms the reliability of current measurements, supporting our findings of the free-falling annular flow's development, which is characterised by an increase in film velocity and a decrease in film thickness at a given inlet condition and within the studied range.

The investigation of entrained droplets has unveiled distinctive profiles for different inlet configurations, particularly in the case of a Tee-junction where the horizontal flow enters the vertical pipe. The turbulence induced by the horizontal flow and flow breakup significantly contributes to a higher entrainment rate in this scenario. Moreover, our findings from the air ventilation tests have revealed that airflow in the central core of the annular flow is not a crucial condition. This discovery underscores the importance of studying droplet entrainment in free-falling flow scenarios, where the presence of airflow in the core could be negligible.

Further research is recommended to investigate wave development and droplet entrainment in scenarios with minimal or no airflow in the core. Improving the understanding of these mechanisms and refining existing correlations will enhance our ability to predict and model such flows accurately. Furthermore, the complex flow behaviour in Tee-junctions warrants further investigation using both qualitative and quantitative methods to advance our knowledge of flow characteristics and droplet entrainment in such configurations.

# 6 Funding

This research was funded by Aliaxis S.A.